\documentclass[12pt]{article}
\usepackage{graphicx}
\begin{document}
\begin{center}
{\Large Is there paradox with infinite space? }\\
\bigskip
{\bf D.H. Coule}\\
\bigskip
 Institute of Cosmology and Gravitation,\\ University of Portsmouth, Mercantile
House, Hampshire Terrace, Portsmouth PO1 2EG.\\
\bigskip

\begin{abstract}

We argue that an infinite universe should not necessarily be
avoided on philosophical grounds. Paradoxes of repeating behaviour
in the infinite, or eternal inflationary, universe can be
alleviated by a realistic definition of differing lives: not
simply permutations of various quantum states. The
super-exponential growth in the rules of a  cellular automata is
used as an example of surpassing the holography bound.

 We also
critically question the notion that our universe could simply be a
simulation in somebody else's computer.
\\


\end{abstract}
\end{center}
\newpage
{\bf Introduction}

There is much discussion about the topology of the universe, most
recently a closed compact Dodecahedral shaped universe  has been
claimed to be consistent with the latest WMAP data [1].
Previously, there was interest in compact open universes - for
general reviews on topology of the universe see [2].

 For a while in cosmology there has been a prejudice against
  infinite universes since
 they apparently give the  paradoxical situation that there must
 also be other identical planets and life to ours. Assuming
 the Copernican principle that our observable part of the universe is not
 unusual, this
is possible in a open or flat FRW universe [3] or if an infinite
number of finite closed universes are created [4] as in the
eternal chaotic universe scenario [5]. One might first wonder if
similar life, beyond the particle horizon in our universe, is more
disconcerting than identical life in other totally disconnected
universes.
 For this reason many
cosmologists prefer the idea of a bound universe to try to prevent
such worrisome ``clones of themselves'' living elsewhere in the
universe. As we later remark this would not anyway be a
satisfactory situation; it implies that all possible human
endeavor is finite but by restricting to a subset we never have to
be aware of this limitation. The only real consideration is that
we should be in a low entropy state as any activity will cause
entropy increase. But while in such a state we are effectively
isolated thermodynamically from the rest of the universe.

Although, we have doubts as to whether certain eternal universe
schemes are viable we wish to argue that this sort of infinity is
not so frightening as often suspected. Because our environment at
our instigation can be extremely  complex there seems little
chance of a life actually repeating the same historic path,
despite an apparent finite number of possible quantum states. We
shall try and argue for a no-repeat condition for life in the
universe. This is independent of global thermodynamical properties
of the universe. We shall use a single cellular automata to
illustrate how the holography bound can be circumvented. Although
the finite number of states of a cellular automata means that it
alone is not sufficient to totally prevent a no-repeat condition,
it does illustrate limitations in how the holography principle has
been  interpreted.

 Some related argument can be made   against
the notion that we are merely the result of somebody's computer
simulation [6]. Such a computer simulation also has problems with
maintaining the notion of free will for life within the
simulation.

Although this debate might seem rather abstract spurious
philosophical notions seem to have been  suggested from such
examples. The infinite universe is said to include somewhere
``every possible action you choose NOT to perform'' [7] . Likewise
in the computer simulation our actions are merely to ``amuse
others'' and so in either case, we are not really under any
compunction to modify our behaviour [8].

{\bf State counting is insufficient}

It is easy to formulate a puzzle with more permutations than
particles in the universe $\sim 10^{80}$. For example a travelling
salesman with $N$  places to visit has $N!$ possible routes. So
$N\sim 100$ routes can surpass this $\sim 10^{80} $ figure. One
can say that the complexity of our environment  is not limited by
this particle number figure, since anyway why should the actual
number of particles vastly distant from us limit what is or isn't
possible for us on earth.

Now if we consider the number of possible quantum states within
the observable universe, which is the essence of the repeat
argument[4,9], we obtain a much larger figure $X\sim
10^{10^{120}}$. This is obtained by applying a holography type
principle to the universe.  According to Rucker [10], who argues
against this sort of reasoning,  E. Wette first obtained $\sim
10^{10^{10}}$ possible space-time states. The actual figure is not
crucial for our argument.  Now is this figure anymore relevant
than the previous number in limiting the complexity of our lives?
Although, a game like Chess or Go does not have this many game
permutations [11] I again can surpass this figure by considering a
more elaborate puzzle: actually by means of cellular automata -
see [12] for a full description.

 For a  block cellular automata
with blocks of size $n$ and allowing $k$ colours there are
$k^{nk^{n}}$ rules or ``games to play'', before we worry about the
number of ways of playing each game [12]. So for say  size $1000$
blocks and ten colours I again can surpass this large figure
apparently imposed by the holography argument. The large number of
quantum states does not constrain the extra patterns that we might
explore even for relatively few atoms involved in such a cellular
automata. Note also that actual information is equal to the
logarithm, to base 2, of the number of possible quantum states. So
the visible universe can only store $\sim 10^{120}$ bits of
information with this many quantum states. This fact will later be
useful in understanding the actual relevance of the holography
bound.

 What possible maximum number of rules, say $Y$, that could be
 explored on a physically realistic computers is unknown but will
in effect be almost infinite, in comparison with the earlier
apparent large number of quantum states i.e. $Y>>X$.
 Such complexity is
probably existing already in nature, for example patterns on cone
shells [12].

  But is this
sufficient complexity to remedy recurrence occurring?  After all
without an infinite grid the number of permutations $Y$ is still
finite and repeats would still eventually occur. But by surpassing
the holography bound there is now no known reason to believe  the
number of possibilities is actually finite. So the resolution of
the  problem is currently unknown. However, I think there is
strong evidence to suppose that the quantum universe should not be
simpler than a classical one.  Recall, if I considered the number
of possible football games and there was no limit to how
accurately I could measure the position of the players and ball I
would  obtain $\aleph_1$ possible games. The same cardinality as
all the points within a 4-dimensional box, whether finite or
infinite sized - see e.g.[10]. So in such a classical universe any
clones do not play exactly the same game regardless of whether the
universe is finite or infinite. Ordinary quantum mechanics does
not alter this since the underlying modes are still continuous,
although such modes are often made of discrete quanta. In many
ways the quantum version is more complicated since Uncertainty
Principles are only limits on complementary measurements while
from Bell's inequality type arguments we know there is not even an
``underlying reality of the universe'' - see e.g. [13]. Indeed,
quantum mechanics allows randomness to remain even when a
measurement has been made. This process itself is rather analogue
in that continuous differing directions can be chosen for the
measurement of say the polarization of photons. As emphasized
recently by Hawking [14] in his so called top-down approach to
cosmology, there is no observer independent history to the
universe. Rather our measurements determine what histories have
dominated in a Feynman path integral for the state of the
universe.

 Whether quantum gravity will eventually show that
 space comes in packets or is continuous should not significantly detract from
 the above reasoning unless quantum mechanics
  is itself simplified by the ultimate quantum gravity theory.
  Presently there still seems many uncertainties in reconciling quantum
mechanical notions with discreteness at the Planck scale cf.
information loss [15]. Incidentally if there is such discreteness
and space-time comes in packets then the infinite universe has at
most size $\aleph_0$ packets.

In conclusion, the actual complexity of our environment although
not precisely known is not simply limited by the number of
particle states within the universe.

{\bf Does Eternal inflation give infinite space?  }

We have previously give a number of possible reasons why the
eternal inflationary mechanism might not hold [16,17] . These
include a) The adaption of the Hawking radiation calculation,
derived from Black hole physics, to de Sitter space not being
entirely correct-see [18] for some differences between the two
cases. This might alter the fluctuation result  required for
galaxy formation. b) Black holes or other topological defects
being produced that reduce the surface gravity and so the
perturbation strength. Especially since the Planck scale might be
altered in Brane models. c) Higher dimensions becoming relevant
that likewise reduce the surface gravity of de Sitter space. In
Brane models the continual presence of a fixed higher dimensional
bulk AdS space could constrain the process cf.[19]. Large
gravitational wave fluctuations from presently inflating regions
might take ``short cuts'' through the bulk space and over produce
matter perturbations in our region of the Brane, or give a large
dark radiation term. d) The required superposition of quantum
modes being prevented by a decoherence mechanism cf.[13]. e) The
quantum fluctuations of matter not automatically being transferred
to the geometric left hand side of Einstein's equation. f) The
imposition of the 2nd law of thermodynamics preventing decrease in
entropy within any horizon volumes- related to the decoherence
mechanism if many other particle modes are present. Other issues
have been emphasized by Turok [20].

We will anyway for now  allow the eternal mechanism the benefit of
the doubt on these sort of issues. However more seriously in this
regard some geodesic incompleteness results for expanding
universes suggests that inflation has only been occurring for a
finite time [21] . So if the initial domain was a closed compact
space only finite numbers of new closed universes can be obtained.
It follows from only so many new Hubble sized domains per unit
time being formed. Actually even with infinite time being
available only $\aleph_0$ universes would be formed. Note that
this argument is similar to one for particle creation numbers in a
steady state universe [10]. There is also a ``new inflationary''
eternal scheme [22] where the universe remains globally  de Sitter
but fluctuations cause local FRW universes to condense out. Again
the finiteness or not depends on the geometry of the background
which is an {\em ab initio} input.

There is yet another variation on this  scheme, involving quantum
tunnelling, where open universes can occur [23]. From within such
a universe the universe appears infinite in size like an open FRW
universe. However, there is some slight of hand with this argument
as observers outside this bubble would still think it is of finite
but growing size. Also  since the time outside the bubble is
limited by the geodesic incompleteness results, and so cannot be
taken to infinity, the full Penrose diagram of the open universe
is not strictly valid - see Fig.(1). We therefore do not think
that the eternal universe mechanism can actually give an infinite
total size universe if the original domain was finite - this claim
suffers from a lack of gauge invariance. It cannot actually change
finite to infinite because of the geodesic incompleteness
restriction. One might start with an initially infinite domain but
then the eternal inflationary mechanism doesn't {\em per se} cause
this infinity. But again for the sake of argument lets still
assume open FRW universe can be formed.
\begin{figure}
\begin{center}
\includegraphics{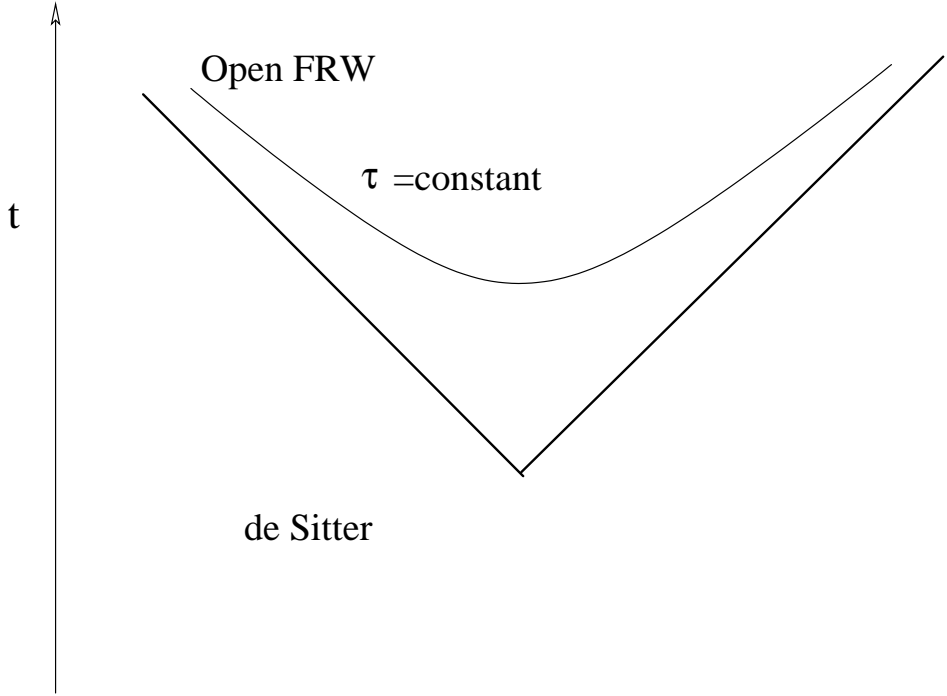}
\caption{ An open FRW universe being created in de Sitter space.
Because the external time $t$ is geodesically incomplete to the
past, a constant time $\tau$  surface within the bubble is only
actually infinite as $t\rightarrow \infty$.}
\end{center}
\end{figure}

{\bf Real life complexity}

 The archetypal infinite universe is
still then just the flat or open FRW universe case [3];  also
present in the pre-big bang model if the initial state is infinite
[24] . By means of a single cellular automata we have  given an
example that allows more rules to explore than are apparently
allowed by an holography argument for the number of quantum
states. This alone pushes repeat lives  vastly further away than
previously expected by quantum state considerations. If cellular
automata are a realistic representation of nature we can speculate
that in our actual lives we are, at least, playing many cellular
automata in parallel. Even this is probably a vast underestimate
since quantum mechanics allows ``non-local'' effects whereas a
cellular automata has only nearest neighbour rules. We therefore
suspect that the actual value of $Y$ will approach $\aleph_0$. If
so then  unless the size of the universe was some higher order
infinity repeats would be absent.

In conclusion, it seem that even if possible identical lifeforms
are present in the infinite universe it can be argued that the
complexity of the games we play or simply ``our lives'' will
fortunately not be being repeated.  In practice we play many
complex games simultaneously and sequentially so the total number
of possible permutations astronomically grows towards what can
only be approximated by countable  infinity $\aleph_0$. There now
seems little difference between an infinite or finite universe in
this regard. A definite proof will require a more complete
understanding of what possibly limits the mathematical complexity
we can interact with, and exactly how future quantum gravity
phenomena should be included.

{\bf Simulation of universe}

Another extreme type of cosmology is that we are simply being
simulated in a computer in somebody else's universe [6]. This is
just the latest version of a long standing idea in cosmology [25].

Bostrom [6] is concerned that if humans don't become extinct they
could start  simulating life on their machines and contribute
excessively to the total number of humans who have ever ``lived''.
We then are most likely if we are a ``typical observer'' to be
actually one of the lives in the machine.

 This problem firstly
concerns deep philosophical issues some of which are closely
related to those considered by Wittgenstein [26,27]. For example,
if nothing around us is then actually real; it is only some sort
of ``activity in a structure'' pretending to be just like  our
universe, how can we then know anything concretely about the true
universe? This argument is related to some in Wittgenstein's {\em
On Certainty} [26] and concerns the difficulty of the skeptic once
all ground rules are done away with. Of course the simulation
exponents want to take notions gleaned from within our ``false''
universe to understand the real universe - but this is a
philosophical impasse.

 Let us think also about simulation. We do not have remotely realistic
  computers to do this but we do
 have some understanding of the  ``computer''  in our heads which
 can simulate or imagine different events. Now a dream in my mind is a
 simulation, but note that the characters in my dream do not have
 true existence as far I am aware.
 There is therefore no essential need for the
 ``people''  within this future simulation to actually exist in the
 sense of self-conscious awareness. What the advocates of this
 scheme are suggesting is therefore a more extreme ``mind
 creation'' device not a simple simulation.

 The connection between a brain state and an actual understanding is also
 more involved than suggested. On a practical ground there are
 estimated to be roughly $\sim 10^{10^{10}}$ possible patterns within
 the brain [28,29]. The notion of keeping track of these fragile states is not
 compatible with  quantum mechanics, since
 observation alone would alter the states.

However, Wittgenstein's {\em ``Philosophical investigations''}[27]
suggests that without ``direct contact'' you cannot gain true
understanding  in your mind. So these created minds must at some
stage have interacted and learnt with real things, that have some
similarity to those in our world. There is a similar reasoned
example by Putnam [30] that if we were simply ``brains in a vat''
being fed impulses  we couldn't refer outside the simulation. We
could never even imagine that we might be all just brains in a
vat. The proponents of the scheme don't object to the universes
being similar to some extent, but we will later show that this
seems incompatible with other considerations.  Although
Wittgenstein is not the last word in philosophy: the notion that
our understanding, or theirs, is strongly grounded in your present
world would place great difficulty in anybody actually simulating
or  understanding what was happening in a totally different world.

Returning to actual computers.  If one takes a cellular automata
[12] then properties like self reproduction can occur. However
this sort of scheme doesn't seem consistent with our notions of
free will. If we instead have free will the simulation cannot
follow a deterministic procedure since it would need to know our
choices beforehand cf. Mackay's argument [31,32] against a
pre-determined universe. This seems especially true in quantum
mechanics where even just the possibility of measurement can be
important cf. delayed choice experiments using photons
gravitationally lensed from  across the universe-see e.g.[13].

For cellular automata,  as mentioned by Barrow [29,32], a
threshold of complexity occurs once self-reference is possible. At
this stage a classical paradox, like ``the Liar's'', can be
described within the system and G\"{o}del like incompleteness is
present. Barrow [33] has suggested errors in the computer
programming could be apparent in our universe since G\"{o}del
incompleteness has been shown to correspond to not knowing the
shortest program to produce a task. However, the physical universe
might not require such an elaborate algebra that contains
incompleteness [32].

 But if we consider the mathematical world
this problem must be apparent. In a simulation nothing is inherent
and so everything has to be put there explicitly.  For example,
there is an interesting problem, beyond current computing
capacity, involving the sign of $Li(n)-\pi(n)$ where $Li(n)$ is
the logarithmic integral function and $\pi(n)$ the number of prime
numbers less than $n$. The value changes sign repeatedly for large
arbitrary numbers -see e.g. [34]. Storing its actual behaviour
seems effectively impossible in any physical system. But if the
computer doing the simulation only works with approximate
algorithms we should be able to overwhelm it, if only by chance we
happen upon an improved algorithm. This is unlike the physical
world which might have completeness.

One might also worry about uncomputable problems in our universe
and how they can be taken care of. For a really astronomical
number consider the Busy Beaver problem with a 12 state Turing
machine [35] and games that might be formulated!

 There are also results [36] that
a computer within our universe cannot processes information as
fast as the universe, or solve every problem we might consider. So
if this result can be extended and applied to the universe in
which the simulation of our universe occurs we either a) have a
much less sophisticated universe than they do or b) the evolution
speed has been  altered to prevent this limitation appearing . The
 constants of nature would apparently
  differ in our universe from theirs ( we must
be running slower). A slower evolving simulation of an inferior
universe  would not be very interesting to the viewers.

 We should also point out that
since the computer cannot totally predict its own universe it is
still rather unprotected. Since the motivation of simulated
universes is to argue against  Doomsday extinction, this computer
still seems exposed to unforseen danger cf.[6]. Especially if
there is some sort of infinite regress of simulations within
simulations. A ``power failure'' in the primary universe would
take them all out. One would anyway expect some sort of
degradation due to simulation i.e. entropy rapidly building up in
the initial actually physical universe.

There must also be some constraint  preventing us developing a
more powerful computer than that doing the simulation - a rather
spurious imposition.

If the universe is really made up of quantum gravity packets of
space cf. loop quantum gravity [37], then these cannot be
compressed further. It violates the quantum rules to define them
more sharply.  Either a) laws of physics are different within the
simulating machine or b) only a subset of the states we think are
in the visible universe are being modelled. Neither is
philosophically appealing.

We can see this problem more directly: recall the universe is made
up of $X$ quantum states which can code   information only of
maximum $\log  X$, see e.g. [38]. But if this is a simulation
within some ``computer '' it would take at least $2^X$ distinct
states to simply note these states within the computer. The
information within the computer is therefore  greater than that
present or ever achievable in the universe. The simulation is
actually massively inefficient. Again one gets the conclusion that
only a vastly inferior, by a two logarithms amount,  universe can
be simulated. This is also related to the underlying randomness
inherent in quantum mechanics needing instead to be accurately
described within the simulating computer, i.e. the so-called
``hidden variables''  are un-hidden in the simulator cf.[13]

 {\bf Conclusions}

By describing a cellular automata with $Y$ possibilities we have
obtained a stronger constraint on the distance that you might
encounter clones of oneself having identical lives. But once the
holography argument is surpassed we see little evidence to further
restrain  the complexity that restricts our lives: on the contrary
quantum mechanics should actually mean the universe is more
complicated than a purely classical universe
 This  might be too subjective for some people since it depends
on us defining structures and games that we interact with to
distinguish different lives.

Recall the archetypal issue is if all possible molecules for life
like DNA have finite variations then repeats happen. Incidentally,
we suspect the way the instructions are implemented allows vastly
more variations depending on chemical concentrations, background
radiation etc, than often supposed. Even then if clones do exist
their differing food sources, environment etc. will cause some
divergences  in their behaviour. However, our main point is that
our games or ``culture''  is so potentially elaborate that no
repeats in life history are realistically possible. Since genetic
twins occur anyway upon earth, and human cloning is nearly
possible, this seems the only realistic form of non-repetition we
could achieve.
  An  implication of this
argument is that whether the universe is finite or infinite
becomes less consequential for the actual complexity of our lives.
Only  a bigger still infinity e.g. $\aleph_1$ of universes, that
is not produced by inflationary expansion, might allow one to
claim repetition is necessarily occurring.  This expanded
complexity of our world can be contrasted with the philosophy
outlined by Wolfram [12] that the universal cellular automata has
the same level of complexity as the whole universe. Although it
might take unreasonable amounts of runtime to calculate the
consequences. Note that unlike Wolfram we are not advocating here
that cellular automata necessarily describe all reality. We have
doubts that, for example, the ``non-local'' effects on quantum
mechanics could adequately be modelled, although some quantum like
automata (complementarity games) have been proposed [39].

 Of course
other reasons might still prefer the  finite case, e.g. having
finite action or simplifying the boundary term for a ``quantum
creation'' calculation. One such finite model is the emergent
universe [40], an update of the Eddington-Lema\^{i}tre universe.
We would just add that one motivation for this model, that it
maximizes the entropy does not seem correct. The later
inflationary state itself allows a higher entropy, and a Black
hole with the same mass produces even more [41,17].

 We also  argued against
the notion that our universe could simply be the inside of a
computer existing within a larger universe. Whether the
hypothetical quantum computer  alters any of these arguments is
uncertain, but the results of Wolpert [36], for example, already
assume arbitrary fast processing. The quantum computer is anyway
likely to be limited by quantum gravity cf.[42]

Let us return to the original motivation for the finite universe.
The argument is that there are only say $N\sim 10^{10^{120}}$
different possible universes of the present visible horizon size.
So if we took a random selection of $N$  closed universes of
similar size to ours,  or if our universe has $N$ times more space
then we should expect repeats to start happening. Many of the
properties of these universes will be constrained by the Laws of
Physics, and initial conditions at the start of the universe. So
many fewer actual alterations are allowed by our own volition.

If the observable universe is growing many more regions are coming
into causal contact so allowing large extra numbers  of possible
quantum states. This increase is likely to dominate over any
possible changes i.e. like moving objects, we might make. So this
sort of reasoning actually ignores the effects of human or any
other life forms intervention - unless they start to develop
galactic scale alterations. The $N$ possible universes are
therefore effectively independent of life which is only a
miniscule perturbation to the actual allowed states. So any actual
universe, say $n$, out of the $N$ possibilities is equally useful
 for our use. Human life which is so paramount to us is just
treated as extreme fine structure in this quantum state number
argument [4,9]. But by the same token it cannot then be applied to
proscribe  life if this is what we are actually interested in
cf.[7]. If the universe is actually accelerating then the argument
can be replaced that we should have only negligible effect upon
the growing entropy of the universe $S=\log W$. In fact the actual
entropy of the visible universe is only $\sim 10^{90}$ -see for
example [38,41], considerably smaller than that given by the
quantum state calculation $\sim 10^{120}$ which is the maximum
possible entropy allowed in the visible universe. It is this  low
entropy that allows us to ``play games'' without restricting their
complexity.

If thermodynamical equilibrium was achieved then the behaviour
would then start to simply ergodically repeat states. So the
repeat arguments  seem actually relevant to a system in
thermodynamic equilibrium where life anyway is not possible. It is
then erroneous  to try and conclude things about life in this
case.
 We have  assumed  a Copernican type principle so that the Laws of
Physics and initial conditions are closely similar. Other work has
considered the possibility of allowing an ensemble of different
Laws or Mathematics which would allow an even greater variety of
possibilities -see also [43,44].

It seems therefore  that as long as our games have negligible
effect upon the universe it doesn't matter whether we are in a
finite or infinite universe. Whether other life forms have exactly
the same culture, music, sport etc. as us depends on other
arguments about how many varieties of such things might be
conceivable. But they are in the finer details of the universe.
Ultimately, the global topology will determine whether ``games''
can continue to be played. For example if the universe continues
accelerating it will reach a stage of heat death with entropy
maximized [17]. But again remaining in a low entropy state and not
the topology of the global universe is the dominant factor.

In conclusion, we don't feel the need to presently exclude
infinite universes necessarily for unwanted philosophical reasons.
Or that infinite production of inflationary universes would mean
``anything in life'' must happen somewhere. Admittedly this
argument might be superseded by future results in quantum gravity
but not by the holography principle {\em per se}. This is
fortunate since the universe regardless of size or state, leaves
us with a blank canvas upon which to impose our own, to us
important, values. In a sense we are disconnected:\footnote {
There is a similar analogy that gravitationally bound systems can
drop out of the expansionary global behaviour of the universe.}
 the infinite universe
isn't some maelstrom dragging everything, including life,  in an
ergodic fashion like atoms in a gas.

 {\bf Acknowledgement}\\ I should like to thank
 David Wolpert for helping explain his results.
\newpage

{\bf References}\\
\begin{enumerate}
\item J.P. Luminet, J.R. Weeks, A. Riazuelo, R. Lehoucq and J.P.
Uzan, Nature 425 (2003) p.593.
\item J.P. Luminet, Phys. Rep. 254 (1995) p
.135.\\
Topology of the universe conference, Cleveland, special issue of
Class. Quant. Grav 15 (1998) p.2529.\\
for a simpler introduction see:\\
W.P. Thurston and J.R. Weeks, Sci. Am. 251(7) (1984) p.108\\
J.P. Luminet,G.D. Starkman and J.R. Weeks, Sci. Am. 280(4) (1999)
p.90.
\item G.F.R. Ellis and G.B. Brundrit, Q. Jl. R. Astr. Soc. 20
(1979) p.37.
\item J. Garriga and A. Vilenkin, Phys. Rev. D. 64 (2002)
p.043511.
\item A.D. Linde, Phys. Lett. B 175 (1986) p.395.
\item N. Bostrom, Philosophical Quarterly, 53 (2003) p.243
\item J. Knobe, K.D. Olum and A. Vilenkin, ``Philosophical
implications of Inflationary cosmology'' preprint physics/0302071.
\item R. Hanson, J. of Evolution and Technology, 7 (2001)
\item M. Tegmark, Parallel universes, in ``Science and Ultimate
reality'' eds. J.D. Barrow, P.C.W. Davies and C. Harper (Cambridge
Press: Cambridge) 2003.\\
M. Tegmark, Sci. Am. ( May, 2003) p. 41.
\item R. Rucker, `` Infinity and the mind'' ( Harvester Press:
Sussex UK) 1982.
\item web site http://Mathworld.wolfram.com/Chess.html
\item S. Wolfram, ``A new kind of science'', ( Wolfram media)
2002.
\item R. Omn{\`e}s, ``The Interpretation of Quantum mechanics'',
(Princeton University Press: Princeton) 1994.
\item S.W. Hawking, lecture at KITP (2003)
\item G. 't Hooft, preprint quant-ph/0212095
\item D.H. Coule , Phys. Rev. D 62 (2000) p.124010.
\item D.H. Coule, Int. J. Mod. Phys. D 12 (2003) p.963.
\item  T.M. Davis, P.C.W. Davies and C.H. Lineweaver, Class. Quant.
Grav.20 (2003) p.2753.
\item K. Kunze, preprint hep-th/0310200.
\item N. Turok, Class. Quant. Grav. 19 (2002) p.3449.
\item A. Borde, A.H. Guth and A. Vilenkin, Phys. Rev. Lett. 90
(2003) p.151301.
\item A. Vilenkin, Phys. Rev. D 27 (1983) p.2848.\\
see also: A.H. Guth, preprint astro-ph/0306275
\item M. Bucher, A.S. Goldhaber and N. Turok, Phys. Rev. D 52
(1995) p.3314.
\item M. Gasperini and G. Veneziano,  Mod. Phys. Lett. A 8 (1993) p.3701.
\item K. Svozil, preprint physics/0305048.
\item L. Wittgenstein, ``On Certainty'', (Oxford: Blackwell) 1969.
\item L. Wittgenstein, ``Philosophical Investigations'' (Oxford:
Blackwell) 1958.
\item R. Rucker, ``Mind Tools'' (Penguin Books: London) 1988.
\item J.D. Barrow, ``The Constants of Nature'', ( Vintage Press:
London) 2003
\item H. Putnam, ``Reason, Truth and History'' (Cambridge
University Press: Cambridge) 1981.
\item D. Mackey, ``The Clockwork image'' (Intervarsity Press:
London) 1974.
\item J.D. Barrow, ``Impossibility'' (Oxford
University Press: Oxford) 1998.
\item J.D. Barrow, New. Scientist, 7th June 2003.
\item K. Devlin, ``Mathematics: The new golden age'' (Penguin
Books: London) 1998.
\item J.L. Casti, ``Five Golden Rules'', (John Wiley: New York)
1996.
\item D.H. Wolpert, Phys. Rev. E 65 (2001) p.016128.
\item A. Ashtekar, preprint math-ph/0202008 and references
therein.
\item S. Lloyd, preprint quant-ph/0110141.
\item K. Svozil, ``Randomness and Undecidability in Physics'',
(World Scientific: Singapore) 1993.
\item G.F.R. Ellis and R. Maartens, preprint gr-qc/0211082.
\item R. Penrose, ``The Emperor's new mind'' (Oxford University
Press: Oxford) 1989.
\item G. 't Hooft, preprint hep-th/0003005.
\item P.C.W. Davies, ``Multiverse or Design'' to appear in
``Universe or Multiverse'' (Cambridge University Press: Cambridge)
ed. B. Carr
\item G.F.R. Ellis, U. Kirchner and W.R. Stoeger, preprint
astro-ph/0305292.

\end{enumerate}
\end{document}